\documentclass{aa}
\usepackage{psfig}
\begin{document}

\title{Wide band X-ray and optical observations of the BL Lac object 
1ES 1959+650 in high state}

\author{G. Tagliaferri\inst{1}
\and M. Ravasio\inst{1}
\and G. Ghisellini\inst{1}
\and F. Tavecchio\inst{1} 
\and P. Giommi\inst{2}
\and E. Massaro\inst{3}
\and R. Nesci\inst{3}
\and G. Tosti\inst{4}
}

\offprints{Tagliaferri}

\institute{INAF-Osservatorio Astronomico di Brera, Via Bianchi 46, I-23807 Merate,
Italy 
\and ASI Science Data Center, Via Galileo Galilei, I-00044 Frascati, Italy
\and Dipartimento di Fisica, Universit\`a La Sapienza, P.le Aldo Moro 2, I-00185
Roma, Italy
\and Dipartimento di Fisica e Osservatorio Astronomico, Universit\`a di Perugia,
Perugia, Italy
}

\date{Received ....; accepted ....}

\abstract{The blazar 1ES\,1959+650 was observed twice by {\it Beppo}SAX
in September 2001 simultaneously with optical observations. We report here
the X-ray data together with the optical, $R_C$ magnitude, light curve since August 1995.
The {\it Beppo}SAX observations were triggered by an active X-ray status of the source.
The X-ray spectra are brighter than the previously published X-ray observations,
although the source was in an even higher state a few months later, as monitored
by the ASM onboard {\it Rossi}XTE, when it was also detected to flare in the TeV band.
Our X-ray spectra are well represented by a continuosly curved model up to 45 keV and are 
interpreted as synchrotron emission, with the peak moving to higher energies.
This is also confirmed
by the slope of the X-ray spectrum which is harder than in previous observations.
Based on our optical and X-ray data, the synchrotron peak turns out to be in the
range 0.1-0.7 keV. We compare our data with non simultaneous radio to TeV data and
model the spectral energy distribution with a homogeneous, one-zone synchrotron 
inverse Compton model. We derive physical parameters that are typical of low power
High Energy peaked Blazar, characterised by a relatively large beaming factor,
low luminosity and absence of external seed photons.
}

\titlerunning{
X-ray and optical observation of the BL Lac object 1ES\,1959+650 in high state}
\authorrunning{Tagliaferri et al.\ }
\maketitle


\section{Introduction}

Blazars constitute the most extreme class of active galactic nuclei: they are 
radio loud AGN characterized by high luminosity, with rapid and high amplitude
variability. They are usually divided into Flat Spectrum Radio Quasars and BL Lac 
objects, according to the presence or not of broad  spectral features.
Another blazar distinguishing  property is the spectral energy distribution 
which is marked by two broad peaks (von Montigny et al. 1995): the first,
extending from the radio to the UV/X--ray band is usually interpreted as synchrotron
emission, while the second, which reaches the $\gamma$--ray band (sometimes 
TeV frequencies) is explained as an inverse Compton component. Based on this,
BL Lacs are further split into two subclasses distinguished by the synchrotron
peak frequency: LBL and HBL, Low and High Energy peaked BL Lacs, respectively
(Padovani \& Giommi 1995). Various models have been proposed to explain the,
still debated, origin of the comptonized photons: they could be the synchrotron
photons themselves (Maraschi, Ghisellini \& Celotti 1992), maybe partly reprocessed 
by the broad line region (Ghisellini \& Madau 1996) or external ambient photons 
emitted by the accretion disk (Dermer \& Shlickeiser 1993) possibly reprocessed
to some extent by the broad line region (Sikora, Begelmann \& Rees 1994) or by
a dusty torus (Bla\.{z}ejowski et al. 2000).

A wide spectral coverage is therefore necessary to understand the physics of 
these objects: accurate simultaneous multiwavelength spectra and light curves 
allow us to constrain the mechanism at work and the geometry of the emitting 
region. HBL in particular have gained much attention in the last decade since
many of them have proved to be TeV emitters: the strongest ones are
the thoroughly investigated Mkn\,421 (Punch et al. 1992) and Mkn\,501 (Quinn et al. 1996). 
Another three sources have then been detected by ground--based
Cherenkov telescopes: 1ES\,2344+514 (Catanese et al. 1998), 
PKS\,2155-304 (Chadwick et al. 1999) and H 1426+428 (Horan et al. 2002).
On the basis of X--ray and radio data, Costamante \& Ghisellini (2002)
selected a catalogue of BL Lac objects which are most likely
to be TeV emitters. One of the most promising objects in this catalogue
is 1ES\,1959+650, as suggested also by Stecker et al. (1996). Indeed, 
this source has now been detected in the TeV band (see below).

1ES\,1959+650 was discovered in the radio band as part of a 4.85 GHz 
survey performed with the 91 m NRAO Green Bank telescope
(Gregory \& Condon 1991; Becker, White \& Edwards 1991).
Subsequently the source was observed also in the optical band
where it displayed large and fast flux variations (e.g. Villata et al. 2000).
It shows a complex structure composed
by an elliptical galaxy (M$_R =-23$) plus a disc and an
absorption dust lane (Heidt et al. 1999).
Its redshift ($z$=0.048) was derived by Perlman et al. (1996)
from a spectrum obtained at the 2.1 m telescope at Kitt Peak.

The first X-ray measurement of 1959+650 was performed by {\it Einstein}-IPC
during the Slew Survey (Elvis et al. 1992). Subsequently, the source was
observed by ROSAT in 1996, by {\it Beppo}SAX in 1997 (Beckmann 2000; 
Beckmann et al. 2002) and finally by USA and RXTE during 2000 (Giebels et al. 2002).
On the basis of a X-ray/radio versus X-ray/optical color--color diagram
the source was classified as a BL Lac object by Schachter et al. (1993).
According to the new schemes, the source belongs to the  HBL class.

In the $\gamma$--ray band, it  was observed in 1995 by EGRET 
on board the CGRO and by the Whipple observatory,
but the source was not detected
and only upper limits could be set (Weekes et al. 1996); 
subsequent EGRET observations allowed Hartmann et al. (1999)
to put it in the third EGRET catalog, with an average flux of
$1.8\times10^{-7}$ photons cm$^{-2}$ s$^{-1}$ at energies above 100 MeV.

During the spring and the summer of 1998, it was observed intensively
with the Utah Seven Telescope Array detector which recorded a 3.9$\sigma$
total significance above 600 GeV. However, during two epochs, the source
was seen above 5$\sigma$ significance (Nishiyama et al. 1999).
These data have been confirmed by the HEGRA and the Whipple teams.
After 94 hours on--source collected in 2000 and 2001, the HEGRA team 
detected the source at $8\%$ of the Crab flux with a 5.4 $\sigma$
significance, while from May to July 2002 the source was detected
in a much brighter state, up to and above the Crab flux (Horns et al. 
2002, Aharonian et al. 2003). Also the Whipple team observed the source 
for $\sim 39$ hours 
between May and July 2002, reporting a mean flux of $0.64\pm0.03$ 
Crab above 600 GeV, with a significance of 20 $\sigma$ (Holder et
al. 2003a,b). Thus, during the May-July 2002 HEGRA and Whipple observations,
1ES\,1959+650 was one of the strongest TeV sources in the sky.

1ES\,1959+650 is therefore one of the most interesting and frequently 
observed sources of recent years. In this paper we present two {\it Beppo}SAX 
ToO observations as part of a project for observing active state blazars
as detected in optical, X--ray or TeV bands. In particular, these 
observations were triggered by the active X-ray state of 1ES\,1959+650
as observed by the {\it Rossi}XTE All Sky Monitor (ASM) and are the final
observations of this successful {\it Beppo}SAX project.

\section{Observations and data reduction}

The Italian-Dutch satellite {\it Beppo}SAX has proved particularly
useful for studying blazars because of its extremely large 
spectral range (0.1-200 keV) which allowed the detection of the 
transition between synchrotron and Compton emission 
(Tagliaferri et al. 2000; Giommi et al. 2000; Ravasio et al. 2002) and 
the interesting comparison of simultaneous soft--X versus hard--X behaviour
(Fossati et al. 2000a; Zhang et al. 2002).

For a detailed description of the mission we refer to
Boella et al. (1997 and references therein). 
In our work we have analyzed the data of different co-aligned 
Narrow Field Instruments aboard the satellite:
the Low Energy Concentrator Spectrometer (LECS; 0.1-10 keV), 
the two (originally three) identical Medium Energy Concentrator 
Spectrometers (MECS; 1.5-10 keV) still operating and the 
passively collimated Phoswich Detector System (PDS; 13-300 keV).

Since 1ES\,1959+650 was observed to be in a high X-ray state by the
{\it Rossi}XTE-ASM during september 2001, we triggered a {\it Beppo}SAX
ToO observation, scheduled to last $5\times10^4$ sec. However, because of
technical problems, it was stopped after $\sim 5\times10^3$ s.
Thus, we were given 
a second opportunity and some days later a new pointing was performed.
In Table \ref{tab1} we show the log of {\it Beppo}SAX observations,
together with exposures and count rates for each instrument.

We based our analysis on linearized and cleaned event files available 
from the online archive 
of the {\it Beppo}SAX Science Data Center (Giommi \& Fiore 1998):
the events from the two operating MECS are merged together 
to improve the photon statistic. Using XSELECT V2.0, we extracted 
light curves and spectra from the event files 
selecting circular regions centered on the source
of 8 and 4 arcmin radii for LECS and MECS respectively, 
as suggested in Fiore et al. (1999). We also extracted events 
from off--source regions to test the background stability.
For the spectral analysis we preferred to use background files accumulated 
from long blank field exposures and available from the SDC public ftp site:
this choice is motivated by the non uniformity of LECS and MECS backgrounds
across the detectors (Fiore et al. 1999; Parmar et al. 1999). In this way
the source and background spectra are extracted from the same detector area.
For the spectral and temporal analysis we used XSPEC 11.0.1
and XRONOS 5.18 packages, respectively.

\begin{table*}[!t!]
\begin{center}
\begin{tabular}{|l|cc|cc|cc|}
\hline
 & \multicolumn{2}{|c|} {\bf{LECS}}  & \multicolumn{2}{|c|} {\bf{MECS}} & \multicolumn{2}{|c|} {\bf{PDS}}  \\
\hline
Date & exposure & count rate$^{\rm a}$ & exposure  & count rate$^{\rm b}$ & exposure & count rate$^{\rm c}$ \\
 & (s) & & (s) & & (s) & \\
\hline
25 September 2001 & $5.8\times10^3$ & $1.12\pm 0.015$ & $7.2\times10^3$ & $1.43\pm 0.016$
       & $ 2.97\times 10^3$ & $0.23 \pm 0.099$ \\
\hline
28-29 September 2001 & $2.6\times10^4$ & $1.23\pm0.008$ & $ 4.8\times10^4$ & $1.75\pm0.007$
       & $ 2.3\times10^4$ & $0.35\pm0.036$\\
\hline
\end{tabular}
\end{center}
\caption{Log of {\it Beppo}SAX observations. $^{\rm a}$ 0.1--10 keV; 
$^{\rm b}$ 1.5--10 keV; $^{\rm c}$ 15--100 keV.}
\label{tab1}
\end{table*}

\section{Spectral analysis}

In order to reproduce the whole X--ray spectrum 
we fitted the data of the LECS, MECS and PDS together: 
to account for the uncertainties in the intercalibration of the instruments 
it is necessary to use constant rescaling factors. The LECS/MECS normalization
acceptable range is [0.67-1], while the PDS/MECS
one is [0.77-0.93] (Fiore et al. 1999). In order to reduce 
the number of fitting parameters and to be consistent 
with previous work, we fixed them at 0.72 and 0.9, respectively. 

\subsection{25 September 2001}

Although the first observation was very short the source was
detected even by the PDS up to 35 keV: the good channels were then 
grouped in 3 bins. We first analyzed LECS + MECS spectra and 
then we added the PDS data: since  for this observation and for 
that of the 28-29 September, the addition of PDS data does not change 
significantly the best fit parameters, in our discussion we will always 
refer to the analysis performed on the full set of LECS--MECS--PDS spectra.

We fitted the data with a power--law or a broken power--law model,
first fixing the absorption parameter N$_{\rm H}$ to
the Galactic value (N$_{\rm H}=1.0\times10^{21}$ cm$^{-2}$
as determined from 21 cm measurements by Dickey \& Lockman (1990))
and then leaving it free: the best fit parameters of each model 
are given in Table \ref{bestfit} together with the flux at 1 keV
and the integrated [2-10] keV flux.
Using the Galactic absorption value we always obtained bad fits:  
large negative residuals toward low frequencies are evident.
With the N$_{\rm H}$ free to vary, instead, we obtained good fits
to the spectrum both with a simple power--law 
( energy index $\alpha = 1.45\pm 0.05$, N$_{\rm H}=2.1\times 10^{21}$ 
cm$^{-2}$; $\chi^2_r/d.o.f.=0.7/32$)
and a steepening  broken power--law model
(N$_{\rm H}=1.8\times 10^{21}$ cm$^{-2}$; $\chi^2_r/d.o.f.=0.6/30$).
The best--fit parameters of the broken power--law model are 
$\alpha_1 = 1.3 \pm 0.1$ and $\alpha_2= 1.5\pm 0.1$ with the break 
located at 2.7 keV.
The $\chi^2$ values do not allow us to distinguish between these spectral
models which are both acceptable. 
Note, however, that in both cases the resulting N$_{\rm H}$ values are
significantly higher than the Galactic value.
 
These results are consistent with previous
X--ray observations of the source. Beckmann et al. (2002)
analyzed the 1997 {\it Beppo}SAX and the  {\it ROSAT}
All Sky Survey data finding a N$_{\rm H}$ value 
in excess of the Galactic value:
more precisely, using a power--law model they found 
N$_{\rm H}=2.55\times 10^{21}$ cm$^{-2}$ for the {\it Beppo}SAX spectrum
and N$_{\rm H}=1.6\times 10^{21}$ cm$^{-2}$ for the {\it ROSAT} spectrum. 
For the {\it Beppo}SAX observation they also fitted a broken power--law
model finding an intermediate value.
This extra absorption required to fit the X-ray data could be due either 
to a real intrinsic absorption at the source or to an artifact due to a 
progressive steepening of the spectrum that can not be reproduced by a 
simple power--law or broken power--law model.
The assumption of an absorption higher than the Galactic 
value would not be necessary if a model which can intrinsically account 
for a progressive steepening of the spectrum is used. 
Therefore, we tried to reproduce the {\it Beppo}SAX spectrum
with the continuously curved model described in Fossati et al. (2000b):
 \begin{equation}
F(E) = K E^{-\alpha_{-\infty}}\big[1+\big(\frac{E}{E_B}\big)^f\big]^{
              (\alpha_{-\infty}-\alpha_{+\infty})/f}
\end{equation}
where $\alpha_{-\infty}$ and $\alpha_{+\infty}$ are the asymptotic
values of the energy indices, while $E_B$ and $f$ are the parameters
that determine the spectral bending. 
We applied this model to the whole LECS--MECS--PDS spectrum
fixing the absorption parameter to the Galactic value.
First we found the value of the parameter $f$ that minimizes the
$\chi^2$ when all the other parameters are left free to vary 
obtainiong $f=1.1$. Then, we repeated the fitting a few times
to derive the spectral indices at several reference energies
(0.5, 1, 10, 35 keV), checking that each time the same values 
for the varying parameters were obtained (see Fossati et al. 
2000b for more details of this procedure).
The data are well fitted, with only the Galactic absorption,
($\chi^2_r/d.o.f.=0.6/31$) by a curve which is flat below 1 keV
and steepens toward higher energies; the model
reaches a slope $\alpha=1$ at $1.2\pm0.1$ keV. In Table \ref{curved} 
we give the best--fit parameters of the model.

A possible inconvenience of this model is that it is so flexible
to adjust the parameters for a wide rage of spectral curvatures and
then it results poorly sensitive to the actual N$_{\rm H}$ value.
Therefore, we also considered another curved model using a logarithmic 
parabola, which provides a reasonable representation of the wide band spectral
distribution for the synchrotron component of blazars:
\begin{equation}
 F(E) = K~ (E/E_1)^{-(a+b~Log(E/E_1))} ~~~.  
\end{equation}
This model has recently been used to fit the optical-X-ray spectra of
various objects (Giommi et al. 2002; Massaro et al. 2003a,b). In our computations we fixed
the $E_1$ value at 1 keV.
With the N$_{\rm H}$ fixed to the Galactic value, this model 
provides a good fit to the data although not as good 
as the previous curved model. A lower $\chi^2$ is obtained if we leave 
the N$_{\rm H}$ free to vary, the resulting best--fit value  
is very similar to the one obtained with the broken power--law model. 
A summary of the best--fit parameters is given in Table \ref{curved} .

\begin{figure*}
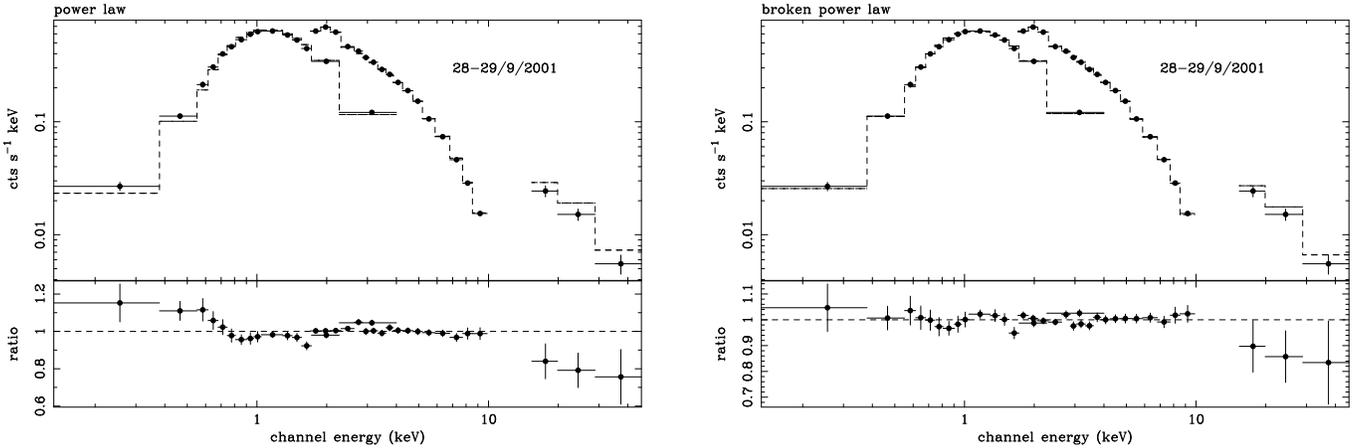

\begin{center}
\hbox to \textwidth{
\centerline{
\vbox {\psfig{figure=fig-1.ps,angle=-90,width=8.5cm}}
\hskip .8cm
\vbox{\psfig{figure=fig-2.ps,angle=-90,width=8.5cm}}
\hfill
}}
\caption{1ES\,1959+650 {\it Beppo}SAX spectrum of 28-29 september 2001.
{\it Left panel}: LECS + MECS + PDS spectrum fitted with a power--law 
plus intrinsic absorption model: 
positive residuals are evident below 0.8 keV, while there are 
negative residuals in the PDS range. {\it Right panel}: LECS + MECS + PDS 
spectrum fitted with a broken power--law plus intrinsic absorption model
provides a better fit, although the PDS data still lie below the model.}
\label{spec2}
\end{center}
\end{figure*}

\subsection{28-29 September 2001}

\begin{figure}
\centerline{
\vbox{
\psfig{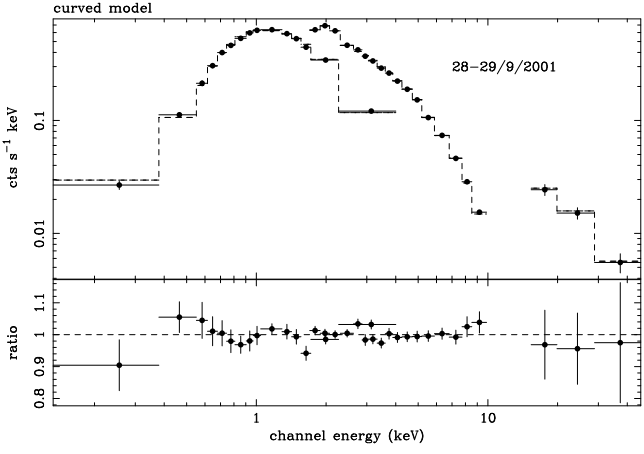}
}}
\caption{1ES\,1959+650 {\it Beppo}SAX spectrum of 28-29 september 2001:
LECS + MECS + PDS spectrum fitted with the curved model. The negative residuals
left by simple and broken power--law models in PDS range disappear.}
\label{curfit}
\end{figure}

The second pointing had an exposure about a factor of 6 longer than
the previous one. Furthermore 1ES\,1959+650 was $\sim$20\% brighter than
in the first observation (see Table 1). We therefore had data with much
higher counts, useful for a more accurate investigation of the spectral
shape. The PDS was able to detect 
the source up to slightly higher frequencies ($\sim 45$ keV) than the 
previous observation. We again grouped the PDS data in 3 bins.
Also in this case we will refer to the analysis performed on LECS+MECS+PDS data
as a whole, repeating the fitting process with the absorption parameter fixed
to the Galactic value or free to vary. 
Spectral fits using power--law or broken power--law models with the 
Galactic N$_{\rm H}$ are not acceptable because of the rather high $\chi^2$
values  (see Table \ref{bestfit}).
With a free N$_{\rm H}$ we again find values that exceed the galactic value
and are consistent with those of the September 25 observation and of Beckmann 
et al. (2002). 
However, in this case the {\it Beppo}SAX spectrum is also badly fitted by 
a single power--law model ($\chi^2_r$/d.o.f.=2.3/32) with the N$_{\rm H}$
free to vary. As can be clearly seen in Fig. \ref{spec2},
this model leaves positive residuals toward low energies and negative 
ones in the PDS range: the spectrum seems to be continuously steepening.
Leaving the absorption parameter free, a broken power--law model with a break
at E$_b \sim 2.6$ keV ($\Delta\alpha\sim 0.26$) provides a better representation
to the data ($\chi^2_r$/d.o.f.=1.0/30), although the negative residuals at
higher energies are still not completely removed (Fig. \ref{spec2}).
Similarly to what was obtained in the analysis of the previous
observation, to fit {\it Beppo}SAX data with a broken power--law model
we need a column density higher than the Galactic value.
We then fitted our data using the curved model of Fossati et al. (2000b).
The data are well fitted with a curvature parameter $f=1.25$ and a 
fixed Galactic absorption value ($\chi^2_r/d.o.f.=1.1/31$). 
In this case the negative residuals left in the PDS range by the 
broken power--law model disappear (see Fig. \ref{curfit}). 
Best-fit parameters are given in Table \ref{curved}.
Similarly to what was observed some days before, the model is hard below 1 keV 
and steepens toward higher energies up to $\alpha_{45 keV} = 1.48$; 
the turning point where the model reaches $\alpha=1$ is shifted to a 
slightly larger frequency ($1.55 \pm 0.05$ keV). 
Finally, we applied the parabolic model, but this time a good
fit could be obtained only with the N$_{\rm H}$ free to vary, finding an
absorption value similar to that obtained with the broken power--law model
(see Table \ref{curved}). 

To summarize our results, on September 2001, 1ES\,1959+650 X--ray spectra 
were convex and could be fitted either with a steepening broken power--law
or a parabolic model with an interstellar absorption exceeding the Galactic
value. Taking into account that optical images of the host galaxy
(see Sect. 5) show that it is crossed by a strong dust lane, it is well 
possible that this additional N$_{\rm H}$ is due to some contribution
within the host galaxy. The higher value of N$_{\rm H}$ derived from the X-ray data
is in line with the E(B-V) values derived from optical observations 
(see section 5). The only model that fits the data with the
N$_{\rm H}$ value fixed at the Galactic value is the curved model proposed
by Fossati et al. (2000b). However, the X-ray spectral shape that one derives
in this case is not very plausible in the overall SED of the source, with a 
synchrotron peak above 1 keV. Thus,
we think that this representation, with the N$_{\rm H}$ fixed at the Galactic 
value, is not realistic. To conclude, our X-ray data imply an absorption higher
than the Galactic value and a peak of the X-ray emission in the 0.1-0.7 keV range.

\begin{table*}
\begin{center}
\begin{tabular}{cccccccc}
\hline
model & N$_{\rm H}$ & $\alpha_1$ & E$_b$ & $\alpha_2$ & F$_{1 keV}$ 
        & F$^*_{2-10 keV}$ & $\chi2_r$/d.o.f.\\
      & ($\times10^{22}$ cm$^{-2}$) & & keV & & ($\mu$Jy) &  &\\
\hline
\hline
\multicolumn{8}{c}{\bf 25 September 2001}\\
\hline
\hline
\multicolumn{8}{c}{LECS + MECS + PDS}\\
\hline
power--law & 0.1 & 1.25 & & & 31.3 & 8.6 & 4.95/33\\
power--law & $0.21\pm 0.03$ & $1.45\pm0.05$ & & & 40.3 & 8.3 & 0.70/32\\
broken power--law  & 0.1 & $0.9^{+0.1}_{-0.5}$ & $1.9^{+0.5}_{-0.7}$ 
          & $1.45\pm0.1$ & 29.5 & 8.3 & 1.20/31\\
broken power--law & $0.18\pm0.03$ & $ 1.3\pm0.1$ & $2.7^{+1.2}_{-0.8}$
          & $1.5\pm0.1$ & 37.2 & 8.3 & 0.55/30\\
\hline
\hline
\multicolumn{8}{c}{\bf 28-29 September 2001}\\
\hline
\hline
\multicolumn{8}{c}{LECS + MECS + PDS}\\
\hline
power--law & 0.1 & 1.15 & & & 33.5 & 10.6 & 24/33\\
power--law & $0.22$ & $1.3$ & & & 42.1 & 10.4 & 2.3/32\\
broken power--law & 0.1 & 0.85 & 2.1 & 1.3 & 30.5 & 10.4 & 2.8/31\\
broken power--law &$0.16^{-0.03}_{+0.02}$ & $1.1\pm0.1$  & $2.6^{+0.5}_{-0.4}$
          & $1.36^{+0.03}_{-0.04}$ & 37.0 & 10.4 & 1.0/30\\
\hline
\end{tabular}
\end{center}
\caption{Best fit model parameters. $^*$  F$_{2-10 keV}$ flux in unit of
$10^{-11}$ erg cm$^{-2}$ s$^{-1}$. }
\label{bestfit}
\end{table*}

\begin{table*}
\begin{center}
\begin{tabular}{ccccccc}
\hline
\multicolumn{7}{c}{LECS + MECS + PDS: curved model}\\
\hline
\multicolumn{7}{c}{\bf 25 September 2001}\\
\hline
$\alpha_{0.5 keV}$ &  $\alpha_{1 keV}$ &  $\alpha_{10 keV}$ &  $\alpha_{35 keV}$ &  
    F$^*_{0.5-2 keV}$ &  F$^*_{2-10 keV}$ 
& $\chi^2_r$/d.o.f \\
\hline
$0.0^{+0.4}_{-0.1}$ & $0.9 \pm 0.1$ & $1.55^{+0.1}_{-0.05}$ & $1.6^{+0.1}_{-0.05}$ 
            & 7.5 & 8.3 
& 0.6/31 \\
\hline
\multicolumn{7}{c}{\bf 28-29 September 2001}\\
\hline
$0.2\pm0.2$  & $0.74\pm 0.03$ & $1.43\pm0.03$  & $1.48^{+0.05}_{-0.04}$ 
                                       & 7.8 & 10.4 
& 1.1/31\\
\hline
\hline
\multicolumn{7}{c}{LECS + MECS + PDS: parabolic model}\\
\hline
\multicolumn{7}{c}{\bf 25 September 2001}\\
N$_{\rm H}$ & $ a $ & $ b $ & F$_{1 keV}$ & F$^*_{0.5-2 keV}$ \\
         & F$^*_{2-10 keV}$ & $\chi^2_r/d.o.f.$   \\
($10^{22}$ cm$^{-2}$) & & & ($\mu$Jy) & & & \\
\hline
$0.17\pm0.06$ & $1.24\pm0.27$ & $0.18\pm0.25$ & 36.9  & 7.49 & 8.26 & 0.63/31  \\
\hline
FIXED & $ 1.925\pm 0.075 $ & $0.465\pm0.1$ & 30.5 & 7.43 & 8.22 & 0.96/32  \\
\hline
\multicolumn{7}{c}{\bf 28-29 September 2001}\\
\hline
$0.16\pm0.03$  & $1.06\pm0.10$ & $0.21\pm0.09$ & 37.0 & 7.82 & 10.4 & 1.18/31\\
\hline
FIXED & 1.83 & 0.4 &  31.7 & 7.82 & 10.4 & 2.4/32  \\
\hline 
\hline
\end{tabular}
\end{center}
\caption{Best fit parameters for the curved and parabolic models of LECS--MECS--PDS 
spectrum. $^*$ fluxes are in units of $10^{-11}$ erg cm$^{-2}$ s$^{-1}$.
For the curved model, the absorption parameter is fixed to the Galactic value
of N$_{\rm H}=1.0\times10^{21}$ cm$^{-2}$; while the best--fit curvature parameters
are $f=1.1$ and $f=1.25$ for the first and the second observation, respectively,
reaching a slope $\alpha = 1$ at $1.2\pm 0.1$ and $1.55\pm 0.05$ keV for the two
spectra.}
\label{curved}
\end{table*}

\subsection{Archival X-ray data}

It is interesting to compare the 2001 {\it Beppo}SAX  spectra to 
those previously measured by other X--ray missions: {\it Einstein}, 
{\it ROSAT} (1996), {\it Beppo}SAX itself (1997), RXTE and USA (2000).
The results of these observations are summarised in Table \ref{hist}.
During {\it Einstein}, 
{\it ROSAT} and the 1997 {\it Beppo}SAX observations, the source had a
flux of F$_{2-10 keV} \sim 10^{-11}$ erg cm$^{-2}$ s$^{-1}$.
In particular, {\it Einstein} measured a 2 keV flux which was $\sim 40\%$
and $\sim 60\%$ higher than the 1997 {\it Beppo}SAX and 1996 {\it ROSAT}
measurements, respectively. 1ES\,1959+650 was monitored by RXTE and  
by USA through the summer and the autumn of 2000. The source was in a higher
state (F$_{2-10 keV} \sim 10^{-10}$ erg cm$^{-2}$ s$^{-1}$), with a flare 
reaching a 2-10 keV flux of $2.3 \times 10^{-10} \
{\rm erg \ cm^{-2} \ s^{-1}}$ (14 November 2000, USA; Giebels et al. 2002).
The source was in a high state also during our observations, similar to 
those measured by RXTE and a little fainter than the flare of November 2000,
monitored by USA.

These X--ray spectra are always soft ($\alpha > 1$) and can be explained as
synchrotron emission. The {\it Beppo}SAX spectra of 2001 are the hardest
measured so far (below 3 keV), as if the synchrotron component were peaking
at higher energies with respect to the other historical observations (see
Discussion).

\begin{table*}[]
\begin{center}
\begin{tabular}{ccccc}
\hline
Instrument & Date & $\alpha$ & F$_{2-10 keV}$ \\
& & & ($ 10^{-11}$ erg cm$^{-2}$ s$^{-1}$) \\
\hline
{\it ROSAT} & 31 March 1996 & $1.76^{+0.44}_{-0.21}$ & 1.04\\
{\it Beppo}SAX & ~4 May 1997 & $1.64\pm0.08$ & 1.29 \\
RXTE & July--September 2000 & 1.38--1.68 & 8.4--14 \\
USA & October--November 2000 & & 4.87--22.6\\ 
{\it Beppo}SAX & 25 September 2001 & 1.31$^a$ & 8.29\\  
{\it Beppo}SAX & 28-29 September 2001 & 1.12$^a$ & 10.42\\
\hline
\end{tabular}
\end{center}
\caption{Historical X--ray observations of 1ES\,1959+650.
$^a$ soft X--ray spectral index.}
\label{hist}
\end{table*}

\section{Temporal analysis}

We examined the observation of September 25 for short time variability,
rebinning the light curves with a resolution of 600 seconds and selecting
only bin intervals with an effective exposure greater than $20\%$
(i.e. all bins that have data for less than 20
bin are rejected).
Because the spectrum seems to steepen at 
$\sim 3$ keV, we analysed the LECS and MECS light curves in two different
energy ranges: [0.3-3] keV LECS and [3-10] keV MECS bands.
During this short period, the observation was interrupted after $\sim 7000$ s, 
the source seemed to be quiet: both LECS [0.3-3] keV and MECS [3-10] keV
light curves are well fitted by a constant model, with a maximum allowed
variability lower than 20\%.

\begin{figure}
\centerline{
\vbox{
\psfig{figure=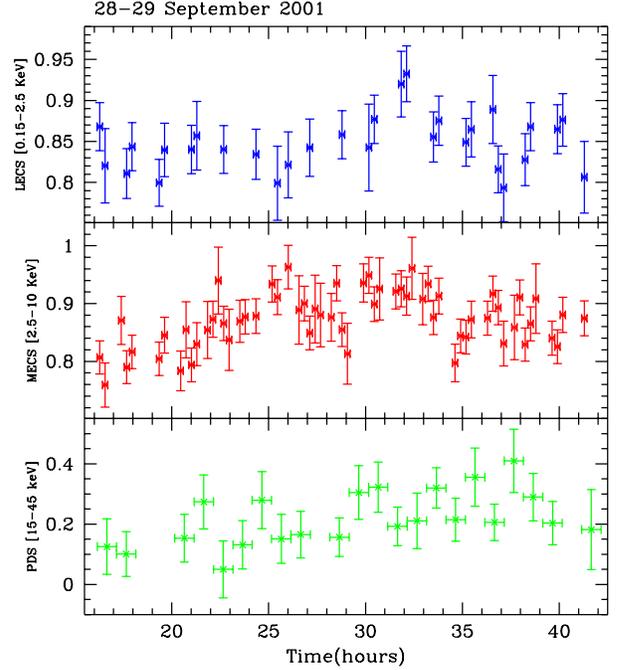,width=9.5cm}
}}
\caption{1ES\,1959+650 28-29 September 2001 light curves.
Top panel: [0.15-2.5] keV LECS $1000 \, s$ binned light curve. 
Mid panel: [2.5-10] keV MECS $1000 \, s$ light curve.
Bottom panel:[15-45] keV PDS $3600 \, s$ binned light curve.
}
\label{2c-ob2}
\end{figure}

The second observation is longer ($\sim 5\times 10^4$ sec) and can 
give us more information. During this run 1ES\,1959+650 was still
displaying a convex spectrum in the LECS-MECS range, with a break
at $\sim 2.5$ keV and a further steepening in the PDS range.
Therefore we again split  our field of analysis extracting
[0.15--2.5] keV LECS and [2.5--10] keV MECS light curves, together
with a PDS curve in the range [15-45] keV (Fig. \ref{2c-ob2}).
For the LECS and MECS data we chose a binning time of 1000 seconds, 
while for the PDS we rebinned the data in 3600 second intervals, due
to lower statistic. In each curve we accepted only bins with at least
$20\%$ of effective exposure time.
Also during this observation {\it Beppo}SAX did not detect very large
variability, with a possible indication of larger variability
above the break. In fact, we get a high probability that the source is
constant below the break for the LECS data, while MECS [2.5-10 keV] and 
PDS light curves are badly fitted by a constant model (constancy probability 
$< 1\%$). This is not surprising since we are observing a steepening 
synchrotron spectra: a little hardening of the synchrotron component 
will cause small flux variations in the proximity of the peak but
larger flux variations as the spectrum steepens at higher energies.
In any case these variations are small and the main conclusion of 
this analysis is that the source did not clearly show brightness
changes during the two {\it Beppo}SAX observations.
Given that the source flux in the second observation increased by only
$\sim$20\% it is likely that it remained approximately stable
at this high level at least from September 25 up to September 29.

\section{Optical observations}

\begin{figure*}
\centerline{
\vbox{\psfig{figure=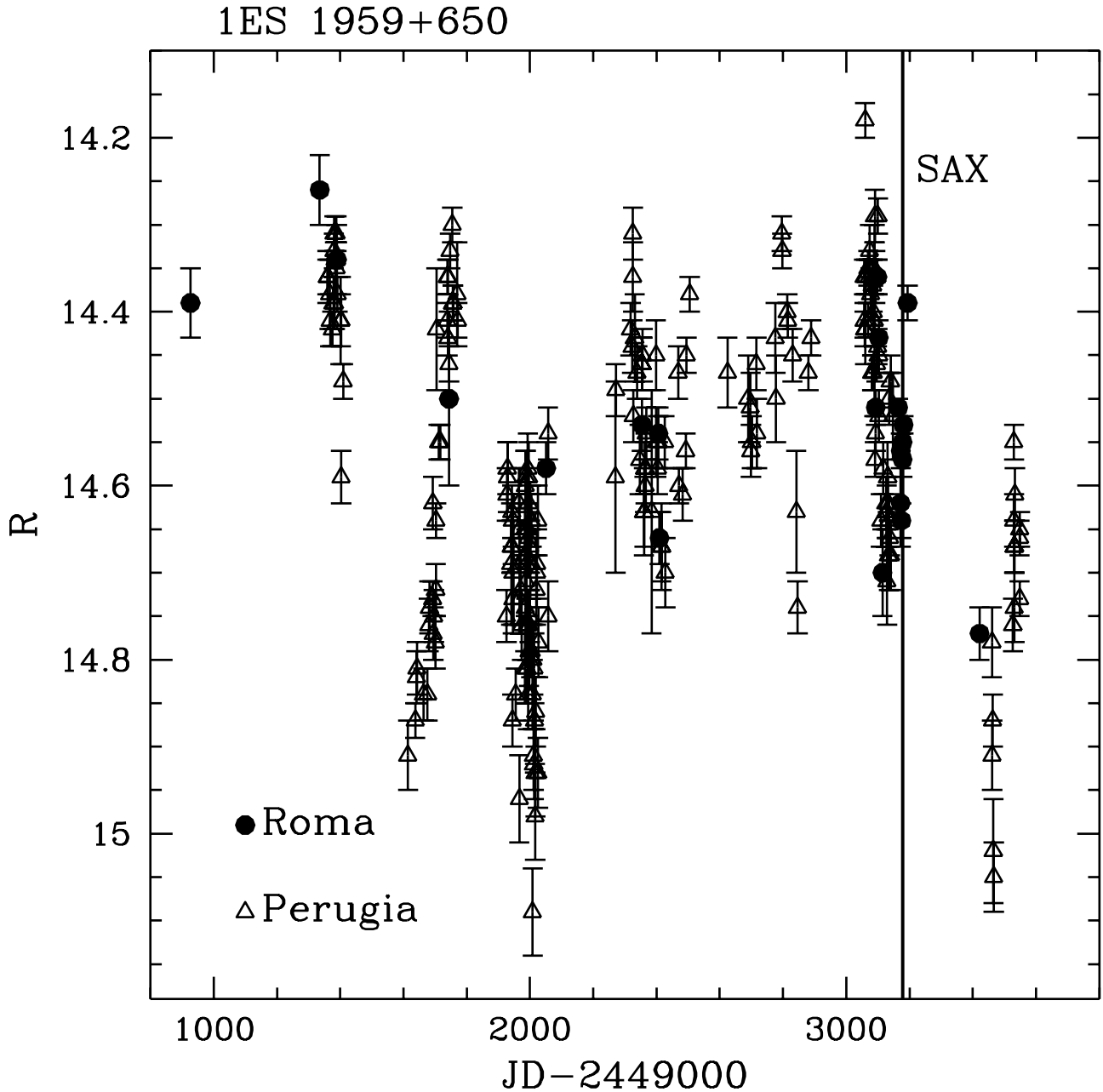}}}
\caption{The optical, $R_C$ magnitude, light curve of 1ES\,1959+650
since August 1995. Open triangles are data from the Perugia observations,
filled circles are from Roma-Vallinfreda. The vertical line indicates the
dates of the {\it Beppo}SAX pointings, activated by an X-Ray trigger.}
\label{optical_lc}
\end{figure*}

We observed 1ES\,1959+650 for about nine hours in the Cousins R band during 
the night of September 28-29 2001 with the 50 cm f/4.5 Newtonian reflector
of the Astronomical Station of Vallinfreda (Rome) (e.g. Maesano et al. 1997).
The reference photometric sequence by Villata
et al. (1998) was used, except for their star 5, which is variable as already
pointed out by Nikolashvili, Kurtanidze and Richter (1999). The source remained
constant around a mean value of R=14.67 within our sensitivity of 0.02 mag
(Maesano et al. 2002).

To measure properly the optical SED of a blazar it is necessary to
subtract from the observed flux values the contribution of the underlying
host galaxy. In the case of 1ES1959+65 the host is detectable even with
the short focal length of our 50 cm telescope.
Heidt et al. (1999) using the 2.5m NOT telescope at La Palma found that a De
Vaucoulers profile ($r_e=9.5''$, R$_{tot}$=14.80) does not give an acceptable
$\chi ^2$ for the fit to the surface brightness of this galaxy, at variance
with other host galaxies of their sample of BL Lacs.
They found a better agreement using a different exponent
(0.41 instead of 0.25), with an effective radius $r_e$=10.3$''$,
a small additional disk component, with $r_e$=4.8$''$ and a total galaxy R
magnitude R$_{tot}$=14.9. 
Furthermore they found a dust lane 1$''$ north of the nucleus.
Scarpa et al. (1999) using HST images confirmed the presence of the dust lane
but found a satisfactory fit to the brightness profile with a De Vaucouleurs
law ($r_e$=5.1$''$, R$_{tot}$=14.92) without the need for an additional disk.
We remark however that the derived contribution of the host galaxy depends
on the seeing conditions: a larger seeing spreads the galaxy image over a
larger apparent radius, so that the galaxy appears fainter within a given
radius. This effect is stronger if the intrinsic $r_e$ is smaller than the
seeing radius, which is not our case, however, whichever estimate of $r_e$ (4.8 or 9.5
arcsec) is assumed.
 
For the purpose of this paper, i.e. to evaluate the contribution of the host
galaxy within our photometric radius, we compute the R magnitude integrating
a De Vaucouleurs profile for both estimates, finding R=15.66 following
Scarpa et al. (1999) and R=15.95 following Heidt et al. (1999) (see Table
\ref{tab_ottica}).
Both values are fainter than the minimum recorded luminosity of the source,
as can be derived from the historical light curve shown in Fig. 
\ref{optical_lc}, built using the data obtained in the course of a monitoring program
of a sample of bright blazars carried out since 1994 by the Roma and Perugia groups.
The telescopes used were the AIT (0.40\,m) of the Perugia Observatory 
(Tosti et al. 1996) and the reflector telescope of Vallinfreda.
The source shows a flux variability between R=14.4 and
R=15.2, without evident periodicity. Some additional photometric
points have been published by Villata et al. (1998), also giving the source
luminosity within the above quoted range.

After subtraction of the galaxy contribution we computed the intrinsic flux 
of the source (in mJy) adopting the extinction curve by Rieke and Lebofsky
(1985) and two possible values for the colour excess:
E(B$-$V)=0.16 (consistent with the Galactic value for the N$_H$, implying
A$_R$=0.43) and E(B$-$V)=0.26, consistent with the N$_H$ value derived from
the X-ray data. The zero magnitude flux value (3.12 Jy at R=0.0) is taken
from Elvis et al. (1994) (see Table \ref{tab_ottica}).

\begin{table*}
\begin{center}
\begin{tabular}{cccccccccc}
\hline
Mag  & Tot. mag & galaxy & Tot & gal & net & Extin. corr. & Intrin. & Extin. corr. & Intrin. \\
band & value    & contr. & mJy & mJy & mJy & E(B-V)=0.16  & mJy     & E(B-V)=0.26  & mJy \\
R & 14.65 & 15.66 & 4.30 & 1.70 & 2.60 & 1.48 & 3.86 & 1.84 & 4.78 \\
R & 14.65 & 15.95 & 4.30 & 1.30 & 3.00 & 1.48 & 4.46 & 1.84 & 5.52 \\
V & 15.11 & 16.43 & 3.19 & 0.94 & 2.25 & 1.59 & 3.58 & 2.12 & 4.77 \\
V & 15.11 & 16.72 & 3.19 & 0.72 & 2.47 & 1.59 & 3.93 & 2.12 & 5.23 \\
\hline
\end{tabular}
\end{center}
\caption{Observed total (blazar+galaxy) R magnitude and derived intrinsic
values for the blazar nucleus. The contribution of the galaxy is calculated
assuming two different profiles (see text).
The V magnitude has not been observed, but it is derived thanks to a 
very constant V-R value observed over the years for this source, see text.}
\label{tab_ottica}
\end{table*}

The V$-$R colour index of the source is remarkably constant in our database,
0.46, with a dispersion of 0.01. Thus, assuming an intrinsic value of 
B$-$V=0.61 for the host galaxy (observed B$-$V=0.77) following Fukugita 
et al. (1995), we can derive the V mag (see Table \ref{tab_ottica}) and a 
spectral energy slope value, in the Log($\nu$)--Log(F$_\nu$) plane, of 0.49 
and 0.83 for the two estimates of the host galaxy magnitude, respectively.

\section{The Spectral Energy Distribution}

We have constructed the SED of 1ES\,1959+650 (shown in Fig. 5) using our
{\it Beppo}SAX and optical data (filled symbols) and data from the literature. 
We also include the previous X-ray spectra measured by ROSAT and 
{\it Beppo}SAX (from Beckmann et al 2002).The TeV spectrum measured 
during a large flare in May 2002 has been recently reported by Aharonian et
al. (2003). The same paper gives the TeV flux observed for the
period 2000-2001. In the SED we assumed that the TeV spectral shape remains
almost unchanged with flux and simply rescaled the spectrum to the
average flux observed during 2000-2001.

The spectral data presented in the previous sections show that the
X--ray spectral distribution of 1ES\,1959+650 is characterised by
a well defined curvature. In the case of the parabolic model, a measure 
of this curvature
is given by the parameter $b$ in Eq.(2). It is interesting to compare 
this result with those of Mkn\,421, whose {\it Beppo}SAX data have been
analysed with the same spectral law (Massaro et al. 2003b).
The typical $b$ values for this source are higher than 0.3 and only 
during the bright phase of spring 2000 are close to 0.2, a value
comparable to that measured in the September 28-29 observation of 
1ES\,1959+650 and equal to 0.21. If the spectral and luminosity trends
of the sources are similar this may be an indication that in the
previous observations, when it was much fainter (see Table 4), the spectral 
curvature could have been greater as also indicated by the steeper spectral slopes.
Furthermore, with this model one can easily derive the energy of the maximum 
in the SED (see Massaro et al. 2003b for details) which is $\sim$0.1 and 
$\sim$0.7 keV in the two observations, respectively.
Similar values are found in the intermediate luminosity states of Mkn\,421,
while during flares the peak ranges between 2.5 and 5 keV. 
We can therefore conclude that the typical peak energies for 1ES\,1959+650 
are around $\sim$0.1--0.3 keV, approaching 1 keV in the X-ray high states.

We have used a homogeneous, one--zone synchrotron inverse Compton 
model to reproduce the SEDs of 1ES\,1959+650.
The model is very similar to the one described in detail in Spada
et al. (2001), it is the ``one--zone'' version of it.
The same model was applied to S5\,0716+714 and OQ\,530
by Tagliaferri et al. (2003);
further details can be found in Ghisellini, Celotti \& Costamante (2002),
where the same model has been applied to low power BL Lacs.
The main assumptions of the model are: 

\begin{itemize}

\item The source is cylindrical, of radius $R$ and
thickness $\Delta R^\prime = R/\Gamma$ (in the comoving frame, where 
$\Gamma$ is the bulk Lorentz factor).

\item The source is assumed to emit an intrinsic luminosity $L^\prime$ 
and to be observed at a viewing angle $\theta$ with respect to the jet axis. 

\item The particle distribution $N(\gamma)$
is assumed to have the slope $n$ [$N(\gamma)\propto \gamma^{-n}$] above the
random Lorentz factor $\gamma_{\rm c}$, for which the radiative (synchrotron 
and inverse Compton) 
cooling time equals $\Delta R^\prime /c$.
The motivation behind this choice is the assumption that 
relativistic particles are injected into the emitting volume 
for a finite time, which we take roughly equal to the light 
crossing time of the shell. This crossing time is roughly equal to the
time needed for two shells to cross, if they have  
bulk Lorentz factor differing by a factor around two.
The electron distribution
is assumed to cut--off abruptly at $\gamma_{\rm max}>\gamma_{\rm c}$.
We then assume that between some $\gamma_{\rm min}$ and $\gamma_{\rm c}$ 
the particle distribution $N(\gamma)\propto \gamma^{-(n-1)}$.
This choice corresponds to the case in which the injected particle distribution
is a power law ($\propto \gamma^{-(n-1)}$)
between $\gamma_{\rm min}$ and $\gamma_{\rm max}$, with 
$\gamma_{\rm min}<\gamma_{\rm c}$.
Below $\gamma_{\rm min}$ we assume $N(\gamma)\propto \gamma^{-1}$.
The value of $\gamma_{\rm max}$ is not crucial, due to the fact
that the electron distribution, for our sources, is steep.
It has been chosen to be a factor $\sim 100$ larger than $\gamma_{\rm min}$.

\end{itemize}

According to our X--ray data, the peak of the synchrotron spectrum
occurs in the 0.1--0.7 keV band.
This motivates the choice of the adopted $\nu_{\rm peak}$
as listed in Table \ref{tab_sed}.
This constrains the values of the input parameters of our model.
Since we determine $\gamma_{\rm peak}$ as the energy of those electrons
that can cool in the injection time (i.e. $\gamma_{\rm peak}=\gamma_{\rm c}$), 
there is a relation between $\nu_{\rm peak}$ and the cooling rate.
If the latter is dominated by synchrotron emission, the location
of $\nu_{\rm peak}$ constrains the value of the magnetic field.
This in turn fixes the relative amount of inverse Compton radiation.

The remaining degree of freedom for the choice of the input parameters
is due to the value of the beaming factor (i.e. $\Gamma$ and the
viewing angle $\theta$), and to the redshift.  Another important input
parameter is the size of the emitting region.  Although the source did
not show fast variability during our {\it Beppo}SAX observations, it
is known to be variable, with a minimum observed variability timescale
of few hours (e.g. Holder et al. 2003a,b; Aharonian et al. 2003).  The
size of the emitting region is therefore constrained to be less than
one light--day (i.e. $R\le ct_{\rm var} \delta/(1+z)$).  Although a
one--zone homogeneous model is forced to use the minimum variability
timescale observed at any band to constrain the size, it is also clear
that this model is a simplification of a scenario which may be more
complex.  Our choice of $R=9\times 10^{15}$ cm and $\delta\sim 18$
corresponds to a minimum variability timescale of $t_{var}\sim 5$
hours.

\begin{figure*}
\psfig{figure=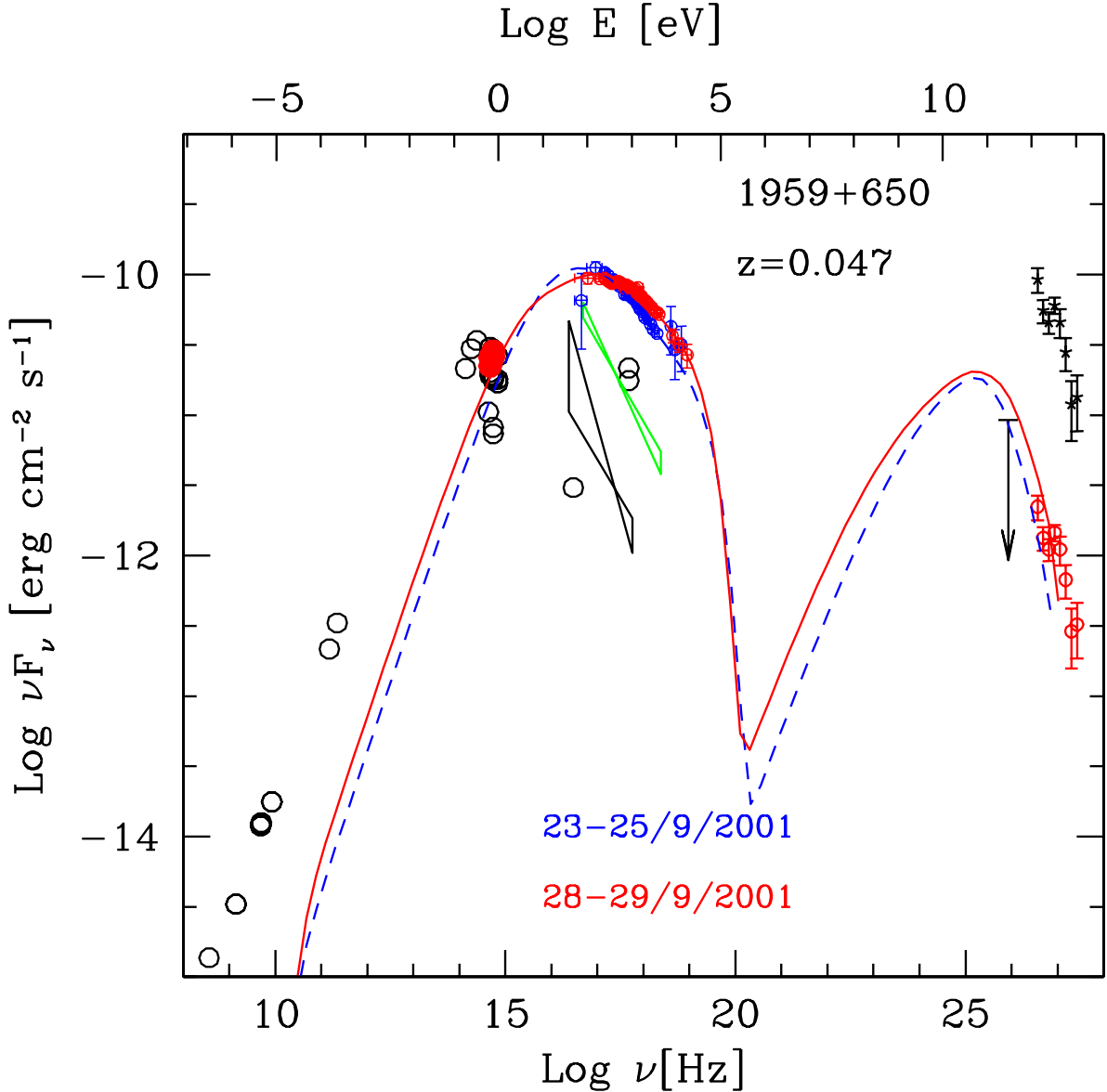,width=18cm}
\vskip -0.5 true cm
\caption{The SED of 1959+650. Non simultaneous data (star symbols)
have been collected from the NASA Extragalagtic Database (NED). The
``butterflies'' in the X-ray region mark the ROSAT (lower) and the
previous {\it Beppo} (upper) spectra (from Beckmann et al. 2000). 
The filled circles represent our optical and X-ray data. The
lines refer to the SSC models for the two {\it Beppo}SAX
observations.  The TeV data (upper symbols) are from Aharonian et
al. (2003) and refer to the flare in May 2002.  For the plotted lower
TeV data we assume no change in spectrum, and we have simply rescaled
the May 2002 data by the amount corresponding to the average flux
observed during the years 2000--2001, as reported by Aharonian et
al. (2003).}
\label{sed}
\end{figure*} 

The input parameters are listed in Table \ref{tab_sed} and
the model fits are shown in Fig. \ref{sed}.
The applied model is aimed at reproducing the spectrum originating 
in a limited part of the jet, thought to be responsible for most of 
the emission. 
This region is necessarily compact, since it must account for the 
fast variability shown by all blazars, especially at high frequencies. 
The radio emission from this compact region is strongly self--absorbed,
and thus the model cannot account for the observed radio flux. 
This explains why the radio data are above the model 
fits in the figure.

\begin{table*}
\begin{center}
\caption{{
Model Input Parameters. 
Column (1): name of the source;
column (2): observation date;
column (3): intrinsic luminosity $L^\prime$; 
column (4): magnetic field $B$;
column (5): size of the emitting region $R$;
column (6): bulk Lorentz factor $\Gamma$; 
column (7): viewing angle $\theta$  (in degrees);
column (8): beaming factor $\delta$; 
column (9): slope of the particle distribution $n$;
column (10): minimum Lorentz factor of the injected electrons $\gamma_{\rm min}$;
column (11): Lorentz factor of the electron emitting at the peaks, $\gamma_{\rm peak}$;
column (12): synchrotron peak frequency $\nu_{\rm peak}$.}
Note that $\gamma_{\rm peak}$ and $\nu_{\rm peak}$ are derived quantities and 
not input parameters.
}
\begin{tabular}{lllllllllll}
\hline
date &$L^\prime$ &$B$ &$R$ &$\Gamma$ &$\theta$ &$\delta$ &$n$  
&$\gamma_{\rm min}$  &$\gamma_{\rm peak}$ &$\nu_{\rm peak}$ \\
        &erg s$^{-1}$ &G  & cm    &     &  &   &    & &     &Hz \\
\hline
23 Sep 2001 &8.0e40  &0.9  &9e15 &14  &3.1 &17.8 &3.8  &1.2e4 &4.0e4 &9.6e16 \\ 
29 Sep 2001 &9.0e40  &0.8  &9e15 &14  &3.1 &17.8 &3.6  &7.0e3 &5.0e4 &1.3e17 \\ 
\hline  
\end{tabular}
\end{center}
\label{tab_sed}
\end{table*}

\section{Discussion and conclusions}

During both {\it Beppo}SAX observations, triggered by an active X-ray status
of the source, we detected X-ray spectra that steepen with 
increasing energies, up to 45 keV. The observed spectra are probably due to
synchrotron emission, with the synchrotron peak moving to higher energies
with respect to previous observations.
1ES\,1959+650 belongs to the class of HBL, being characterized by 
a synchrotron peak in the soft X--ray range.
This is confirmed by our {\it Beppo}SAX observations,
which caught the source in a high X--ray state.
The slope of the X--ray spectrum is harder than during
previous X--ray observation (see Fig. \ref{sed}),
suggesting a ``harder when brighter" behavior,
common to many other blazars, at least at frequencies
above the synchrotron peak (indications exist that the
same behavior is shown also at frequencies above the
inverse Compton peak, see e.g. 3C\,279, Ballo et al. 2002; PKS\,0528+134, Ghisellini et al. 1999).

We do not know yet, however, if 1ES\,1959+650, at still
higher X--ray (2--10 keV) fluxes, is similar
to the flaring state of Mkn\,501 (during summer 1997)
with a flat ($\alpha_x<1$) X--ray spectrum peaking at
hundreds of keV, or resembles instead the X-ray spectrum
of Mkn\,421 which has been always observed with $\alpha_x>1$.
The physical parameters derived by applying our one--zone
SSC model are typical of all low power HBL objects, 
characterized by a relatively large beaming factor,
low luminosity and absence of external seed photons
testified also by the very weak (if any) broad emission lines.
One can compare the parameters derived in this paper with those
derived by a sample of BL Lac objects in Ghisellini, Celotti \& 
Costamante (2002) to see that all sources in this class are
characterized by similar parameters.
These are the sources which are among the best candidates to be
strong TeV emitters (see Costamante \& Ghisellini 2002), 
and in fact 1ES\,1959+650 recently
has been detected at TeV energies at levels above the Crab
by the WHIPPLE and HEGRA Cherenkov telescopes 
(Holder et al. 2003a, Aharonian et al. 2003).
At the time of the TeV flares the X--ray flux was larger than during our
{\it Beppo}SAX observations, in fact it was in the brightest X-ray
state of the source in the last seven years, as monitored by ASM
onboard {\it Rossi}XTE (Holder et al. 2003b). This confirms the strong 
correlation between X--ray and TeV emission, in turn confirming that the
same population of electrons are emitting in both bands.
However, one very rapid TeV flare (doubling timescale of 7 hours)
was detected by WHIPPLE without a corresponding brightening of the 
X-ray flux (Holder et al. 2003b). This is clearly difficult to explain
with one-zone SSC model. If more examples of this intriguing flare
episode will be detected in this and other sources, then some extra
complexity in the model will be necessary.
 
This is the second source in our {\it Beppo}SAX ToO program that has been 
observed because it was in an active X-ray state, the other one being PKS2005-489
(Tagliaferri et al. 2001). In both cases, only the synchrotron component 
was observed with the synchrotron emission peaking at 1-2 keV and
no fast variability was detected. For six other sources the trigger
came from the optical band. Thus, we have probably been biased
toward sources that have an higher optical variability.
These should be the blazars which have the synchrotron peak 
in the IR-optical band and these are of course essentially LBL
or intermediate blazars. For all of them we detected either only 
the Compton component or both the synchrotron and Compton components
(see Tagliaferri et al. 2002). Moreover in three cases we detected very fast 
variability in the X-ray band, but only for the synchrotron component
(ON231: Tagliaferri et al. 2000; BL Lac: Ravasio et al. 2002; S5 0716+714:
Tagliaferri et al. 2003). This can be interpreted as the
presence in the X--ray band of a Compton component (slowly variable on
time scale of months), and the tail of a synchrotron component with 
fast and the erratic variability.

Thus, we can conclude that, although our sample of {\it Beppo}SAX ToO observations 
is limited, the behaviour of LBL-intermediate blazars in the X-ray band is 
different than that of HBL, when they are both in a high state of activity.

\begin{acknowledgements}

This research was financially supported by the Italian Space Agency and MIUR.
This research made use of the NASA/IPAC Extragalactic Database (NED) which is
operated by the Jet Propulsion Laboratory, Caltech, under contract with NASA.
\end{acknowledgements}

\end{document}